\documentstyle{laa}
\begin{document}
\thesaurus{07.33.1,07.28.1,07.34.1}
\title{The thick disc of the Galaxy: Sequel of a merging event}
\author{Annie C. Robin \inst{1,2} \and Misha Haywood \inst{1} \and
Michel Cr\'ez\'e \inst{2} \and Devendra
K. Ojha \inst{1,2} \and Olivier Bienaym\'e \inst{2}}

\offprints{A.C. Robin}
\institute{Observatoire de Besan\c{c}on, BP1615, F-25010 Besan\c{c}on cedex,
France  \and CNRS URA1280 Observatoire de Strasbourg, 11 rue de
l'Universit\'e, F-67000 Strasbourg, France}
\date{Received date: 10-mar-1995; accepted date: 25-apr-1995}
\maketitle
\renewcommand{\deg}{^{\circ}}
\begin{abstract}
Accurate characterization of thick disc properties from recent
kinematic and photometric surveys provides converging evidences that
this intermediate population is a sequel of the violent heating of
early disc populations by a merging satellite galaxy.

The thick disc population is revisited under the light of new data in
a number of galactic sample fields. Various thick disc hypotheses are
fitted to observational data through a maximum likelihood
technique. The resulting characteristics of the thick disc are the
following : a scale height
of 760 $\pm$ 50 pc, with a local density of 5.6 $\pm$ 1 \% of the thin
disc. The scale length is constrained to be 2.8 $\pm$ 0.8 kpc, well in
agreement with the disc scale length (2.5 $\pm$ 0.3 kpc). The mean
metallicity of the thick disc is found to
be -0.7 $\pm$ 0.2 dex, with no {significant} metallicity gradients.

These photometric constraints in combination with kinematic data give
new constraints on the thick disc formation. We show that thick disc
characteristics are {hardly compatible with a top-down
formation scenario}
but fully compatible with a violent merging event arising at
the early thin disc life time as described by Quinn,
Hernquist \& Fullagar (1993).

\end{abstract}

\section{Introduction}

After a decade-long controversy about the existence or non-existence of
a thick disc in the Galaxy, data accumulate to support the existence
of such a component. Many observations
show that this population is distinct from the disc and the halo (see
Reid \& Majewski, 1993, hereafter RM, for a review). However the
formation of this component is still an open question. In the absence
of a
complete description of the population,
it is almost impossible to design a
probable scenario. Yet understanding this population is a necessary
step towards understanding the galaxy formation, halo collapse, disc
collapse and disc dynamical and chemical evolution.

The reason why the characteristics remain controversial is related to
the fact that its members cannot be easily recognized from the disc or
the halo members in most
observable distributions, such as magnitude - colour
diagrams, or even in reduced proper motion diagrams, specially if the
photometry is not accurate enough. More discriminating data, like
metallicity and radial velocity distributions, hardly constrain the
scenario because they generally concern a too small number of stars
on a small area of the sky.  The determination of the thick disc structural
parameters requires large star samples in various directions
well distributed in longitude and latitude. As a result
the density law, the local density and the metallicity of the thick
disc still remain uncertain.

While it is generally
assumed that the density law can be modelled by a double
exponential, the scale length is unknown and
scale height measurements range run between 700 pc (Yamagata \&
Yoshii, 1992) and 1500 pc (Reid \& Majewski, 1993 ; see this reference
for a review of recent determinations). In most galactic models the estimated
thick disc scale height seems to be correlated with the assumed
disc scale height, since in B-V counts the two populations are not well
separated. U-B colour distributions give a better separation between
the two populations due to blanketing effect, but they are scarcely
available due to the difficulty of getting good calibrations in the U band.

The local
density of this population has hardly been measured directly in the solar
neighbourhood. Several attempts have been made using high proper motion
stars (Sandage \& Fouts, 1987a, Casertano et al., 1990). They do not
give an accurate measurement ($\rho_{0} > 2 \%$ for the former, $
10 \pm 5$ \% for the latter). This method is more reliable for halo
stars because they are easily distinguished from disc stars by their
kinematics.

The local density of the intermediate
population can be deduced from remote star counts but such estimates
are correlated with the
scale height. Large local density estimates are obtained in
combination with short scale height,
while small local density are associated with
large scale heights (e.g. RM found 2.0-2.5 \%
with a scale height of 1400 pc). Estimates range between
1\% of the disc density to more than 10\%.  RM
argue from the analysis of their own data towards the North Galactic
Pole that the thick disc scale height cannot be measured
independently from its local normalization. However combining data at
different latitudes and longitudes this
apparent degeneracy can be removed.

The metallicity distribution of the thick disc has never been
accurately measured
because of the lack of complete samples of thick disc stars free
from observational bias. It is generally admitted
that its mean metallicity is compatible with the one of the disc
globular clusters (-0.6 to -0.7) but could extend from -0.5 (the
supposed limit for the old disc) to -1.5 (Morrison et al., 1990).

Most investigations based on photometric star counts try to derive
thick disc characteristics through direct inference from data in one
field. As a result, since broad band photometry is not sufficient to
unambiguously discriminate thick disc stars from halo and thin disc
ones, estimates of the properties of this component are biased by
population confusion. In contrast in a fully synthetic approach
fitting model predictions to data in many fields, the discrimination
is based on the characteristics of observed distributions. Then model
components are resolved under strict control.

The Besan\c{c}on model of stellar population
synthesis has been developed in order to study the galactic structure
and to understand the formation and evolution of the Galaxy (Robin \&
Cr\'ez\'e, 1986, Bienaym\'e et al., 1987, Haywood, 1994). In
parallel to this approach, an observational programme has been started
in collaboration between the french observatories of Besan\c{c}on,
Strasbourg, Paris and the Uttar Pradesh State Observatory, featuring
photometric and proper motion star counts in several
directions.  These data combine Schmidt plate UBV photometry and
proper motions up to magnitude 18 or 19 in V and complementary deep
CCD data in few sub-fields. A description of the programme may be found
in Robin et al. (1993)

In this paper we describe an attempt to solve for the main
characteristics of the thick disc population (scale height, scale
length, local density, metallicity) using photometric star counts in a
number of directions and the model of population synthesis. We present
in section 2 the data set used in this analysis. In section 3 we
describe the characteristics of the model simulated data. The
statistical tests applied to photometric data to obtain the structural
parameters of the thick disc are described in section 4 and
5. Implications for the thick disc nature and origin, in
combination with other kinematic data, are discussed in
section 6.

\section{Star counts}

In addition to data from our own survey, star count data from a number
of other investigations have been compiled and entered in the global
match. However the investigation has been limited to data sets
including at least the V and B band.

\subsection{Photometric catalogues}

The Besan\c{c}on photometric and astrometric sample survey supplies
photometric data samples with an accuracy better than 0.1 magnitude in
UBV in 5 fields. Among those, here we use the fields at intermediate
latitude on the galactic meridian and one field toward the galactic
pole :

\begin{itemize}
\item A field near the globular cluster M5 (l=2.7$\deg$, b=47.4$\deg$)
(Bienaym\'e et al. 1992,
Ojha et al., 1994b, hereafter called ``Galactic Center Intermediate
Latitude'' GCIL) which covers 15.5 square degrees and is complete to
18 in V, 19 in B and 18 in U.

\item A 7.13 square degree field at l=167.5$\deg$ and b=47.4$\deg$
(Ojha et al., 1994a) complete to 18.5 in V, 20 in B. We call it
Anticentre Intermediate Latitude, ACIL.

\item A field close to the north galactic pole, near the globular
cluster M3 (Soubiran (1992)) (l=50$\deg$, b=80$\deg$) covering 20.26
square degrees and complete to about 17.5

\item A field in the direction of antirotation at intermediate
latitude (l=277.8$\deg$, b=46.7$\deg$) covering 20.84 square degrees
and complete to 18.5 in V, 20 in B, 18.5 in U.

\end{itemize}

\noindent Other star counts surveys suitable for our analysis include :

\begin{itemize}
\item The Chiu's (1980) survey in three selected areas (SA57 near
the north galactic pole, SA68 (l=111 $\deg$, b=-46 $\deg$), SA51
(l=189$\deg$, b=21$\deg$). Each field covers 0.1 square degree and
reaches magnitude 20 in V and 21 in B.
\item The Gilmore, Reid \& Hewett (1985) survey concern
two fields: the SGP on 11.5 square degrees and a field in Aquarius
(l=37$\deg$, b=-51$\deg$) on 18 square degrees. Both fields are
complete to about 18 in B and V.
\item The Basle Halo Programme (hereafter BHP) contains a number of
fields in three colour photometry on small areas of sky, of the order
of 1 to 2 square degrees. Most of them are in the RGU system and
require a conversion to the Johnson system. Few fields of the BHP are
also available in UBV : SA57 by Fenkart \& Esin-Yilmaz (1985) and
Spaenhauer (1989 and private communication), and SA54 (l=179$\deg$,
b=51$\deg$) by Fenkart \& Esin-Yilmaz (1983) and are used in the
present study.

\item Stobie \& Ishida (1987) produced UBV star counts towards the NGP
from Schmidt plates on 21.46 square degrees. The data are complete to
about 18 in V.

\item Yamagata \& Yoshii (1992) field is in SA54 (l=200$\deg$,
b=59$\deg$). It is complete up to magnitude 18 in V.
\item Ratnatunga (1984) explored 3 fields (SA127 (l=272$\deg$,
b=-38.6$\deg$), SA141 near the south galactic pole, SA189 (l=277$\deg$,
b=-50$\deg$) in a photo\-metric and spectroscopic survey
between 9 and 14 in V on areas of about 20 square degrees.
\item Friel and Cudworth (1986) observed two fields in a photometric and
spectroscopic survey :
Serpens (l=37$\deg$ , b=+51$\deg$) and SA94 (l=175$\deg$,
b=-49$\deg$). Their BV
sample is limited to 14 in V and covers only 4 square degrees.
\item Deeper counts have been obtained on smaller areas from prime
focus plates in J
and F between magnitude 20 and 22. Kron (1980) and Koo and Kron (1982)
observed SA57 and SA68 on 0.1 square degree. At very faint magnitudes the
star-galaxy separation is doubtful on photographic plates. Trevese \&
Kron (1988) looked for variabilities in the same field in several
colours (U,J,F,N) and verified the star-galaxy
separation from the Koo \& Kron sample. The resulting sample is
probably free from extragalactic objects.
\item Majewski (1992) obtained new star counts in SA57 from prime
focus plates in several colours and proper motions. He surveyed an
area of about 0.3 square degree and is complete to V=22.5.
\end{itemize}

\subsection{Data accuracy}
When comparing colour distributions in neighbouring fields from
different authors, star density discrepancies are sometimes larger
than the Poisson noise and larger than photometric random errors, as
estimated by the authors. In figure~1 we compare B-V histograms from
these wide field surveys in the polar caps (SI, GRH and
Soubiran). Disagreements appear well before the limiting magnitude and
specially in the wings of the distributions. In U-B the problem is
even more acute. Figure~2 shows U-B histograms in different V
magnitude bins at the pole from different authors. There is no
discrepancies between Spaenhauer and Fenkart \& Esin-Yilmaz data but
Stobie \& Ishida data appear systematically shifted by 0.1 to 0.2
magnitudes. The same problem appears between Fenkart \& Esin-Yilmaz
and Yamagata \& Yoshii colour distribution towards SA54.  The
discrepancies could be due to true differences from field to
field. This would be a signature of local inhomogeneities of the
Galaxy. If these differences are real we should use a mean of these
distributions so as to get the mean properties of the
populations. However, the existence, in some cases, of a disagreement
appearing only in U-B and not in B-V indicates that it is most likely
a problem of photometric systematic errors. It is well known that
calibration of Schmidt plates leads easily to systematic errors if
standard stars are too scarce, or not well distributed in magnitude
and in position over the plate. This is often the case in the U band
and the non-linearity of the calibration leads to systematic
deviations at magnitudes at which no standards are available. It
should be noted that Spaenhauer and Fenkart \& Esin-Yilmaz use the
same plate-emulsion combination and the same scanning machine. They
obtain the same realization of the UBV system,
while this is not the case for the other data sets. As a conclusion,
one need to take into account, in the analysis, star count errors
larger than the Poisson noise.
Discrepancies as large as 0.1 mag. in U-B can be observed between
two different implementations of the same photometric
system. Furthermore while each group claims reliability and
completeness of their photometry down to magnitude 18, obvious
discrepancies cast doubt on any interpretation beyond m$_V=17$.

\begin{figure}
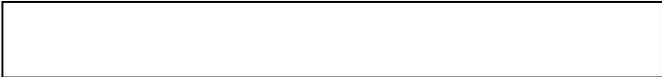

\picplace{1 cm}
\caption[]{B-V distribution towards galactic poles in V magnitude bins
from Stobie \& Ishida (1987, squares), Soubiran (1992, diamonds)
and Gilmore et al. (1985, crosses). Error bars are 1 sigma
Poisson noise.}
\end{figure}

\begin{figure}
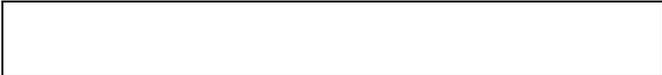

\picplace{1 cm}
\caption[]{U-B distribution towards galactic poles in V magnitude bins from
Stobie \& Ishida (1987, crosses), Fenkart \& Esin-Yilmaz (1985,
squares), Spaenhauer
(1989,  diamonds). Error bars are 1 sigma Poisson noise.}
\end{figure}

\section{Modelling aspects}

The Besan\c{c}on model of stellar population synthesis has been
developed for about 10 years as an attempt to put together all
constraints (theoretical and observational) about galactic evolution
in order to obtain a consistent scenario of galaxy
evolution. The steps of this approach have been the following : Robin \&
Cr\'ez\'e (1986) built a population synthesis model using an
evolutionary scheme from Rocca-Volmerange et al. (1981) suitable for external
disc galaxies. Bienaym\'e et al. (1987) used dynamical constraints
(Boltzmann and Poisson equations and the observed rotation curve) to
determine the scale heights of the disc, according to the potential of
a self-consistent mass model. The kinematical parameters are
function of age (Robin \& Oblak, 1989). Then, in order to investigate the
Initial Mass Function and the Star Formation history in the galactic
disc, Haywood (1994) improved the evolution computation using
the most recent evolutionary tracks and obtained strong constraints on
the slope of the IMF, the history of the SFR in the past and on the
age-vertical velocity dispersion relation, i.e. on the process of
orbit diffusion in the disc. The model we use in this analysis is the
best fit model determined by Haywood (1994). The evolution of
the disc is described by a constant star formation rate, an
age-velocity dispersion relation rising to 21 km s$^{-1}$ for oldest
disc stars and a three-slope Initial Mass Function. Evolutionary
tracks are from Schaller et al. (1992) and Vandenberg (private
communication).

The model, initially limited to UBV photometry, has been extended
to red and near infrared bands (R I J H K L), (Robin, 1993)
and to IRAS 12 microns and 25 microns bands
(Guglielmo \& Robin, 1993).

Thin disc metallicities are a function of age, following Twarog
(1980). A natural metallicity dispersion at each age ranging from 0.12
dex at 1 Gyr to 0.20 dex at 10 Gyr. The oldest thin disc stars have
metallicities of -0.37 $\pm$ 0.2.  Since old stars have also larger
scale heights than young ones, this creates a vertical metallicity
gradient in the disc of about -0.15 dex kpc$^{-1}$ in the case of our
metallicity-age-scale height relation. This vertical disc gradient is
not a free parameter in the model. It is given by the age-metallicity
and age-scale height relations. {More recent age-metallicity relation
(AMR) from Edvarsson et al. (1993) give a slightly higher metallicity
for old disc stars and a smaller age-metallicity dependence. They find
a mean metallicity of -0.25 dex for disc stars older than 6.3 Gyr at
the solar position. But according to the large dispersion in
metallicity for a given age (about 0.2 dex) they also claim that their
AMR is consistent with Twarog (1980) and Meusinger et
al. (1991). Actually the results of the present paper on the thick
disc characteristics do not depend on the assumed AMR for the disc
because colour bins used in the model fitting are widely dominated by
the thick disc population.}

The blanketing vector calculations have been computed using the model
atmospheres from Buser \& Kurucz (1992). Since the blanketing
correction is a non-linear function of the metallicity, even for small
metal deficiencies, we compute for each luminosity class a blanketing
correction in U-B and B-V as a polynomial function of [Fe/H] and
B-V. These functions were computed in the domain where the atmosphere
models are available, that is from spectral types F to K. For later
type stars we extrapolated, assuming that the blanketing effect
becomes negligible for red dwarfs at about B-V=1.5, as can be seen in
models from Bergbush \& Vandenberg (1992). No U-B star counts are
presently available for metal poor M dwarfs in order to check this
assumption. For early type stars with B-V $<$ 0.1 the blanketing
correction is supposed to be 0. For 0.1 $<$ B-V $<$ 0.54 (for which
reliable atmosphere models are not available) we interpolate the
blanketing between the values at 0. and 0.54. This has no effect on this
study where late populations dominate.

For thick disc and halo populations we adopt the M$_{V}$ vs B-V relation
given by Bergbush \& Vandenberg (1992) and we apply the Buser \&
Kurucz models to compute the U-B colours for a set of 4 metallicities
( -0.65, -1.3, -1.6 and -2.2). Intermediate metallicity models are
obtained by linear interpolation. Figure~3 shows the resulting
colour-colour diagrams at different metallicities. The thick disc
and halo luminosity functions are those of 47Tuc and M3 respectively.

\begin{figure*}
\picplace{1 cm}
\caption[]{(U-B,B-V) diagrams for giants and dwarfs at metallicities 0,
-0.5, -1.0, -1.5 and -2.0 which are used for thick disc and halo
populations.}
\end{figure*}

\section{Fitting Method}

\subsection{Maximum likelihood}

In order to determine the structural parameters of the thick disc, we
produce a grid of thick disc models, with different values of the
local density, scale height and scale length.  For each thick disc
model we simulate catalogues of data similar to the observed data
sets, including photometric errors according to the authors. For a
better statistics and to minimize the Poisson noise in model
simulations, small field catalogues (areas smaller than 5 square
degrees) are simulated ten times. Then we compute the likelihood of
the observed catalogues to be a realization of each model. The
likelihood is computed as described in Bienaym\'e et al. (1987,
appendix C) :

Let $q_{i}$ be the number of stars predicted by the model in bin $i$
and
$f_{i}$ be the  observed number. In case the deviations of  $f_{i}$'s
with
respect to  $q_{i}$ just reflect random fluctuations in the counts,
each $f_{i}$ would be a Poisson variate with mean  $q_{i}$. Then
the probability that $f_{i}$ be observed is :

\begin{equation}
dP{_i} =   \frac{q_{i}^{f_{i}}}{f_{i}!}  exp ( - q_{i})
\end{equation}

Then the likelihood of a set of $q_{i}$'s
given the relevant $f_{i}$ is :
\begin{equation}
L = ln \sum dP_{i} = \sum_{i} (- q_{i} + f_{i} \ln q_{i} - \ln
f_{i}!)
\end{equation}

In search for the models that maximize $L$ it is convenient to use the
reduced form :

\begin{equation}
L - L_{0} = \sum_{i} f_{i} (1-\frac{q_{i}}{f_{i}} +
ln\frac{q_{i}}{f_{i}})
\end{equation}

\subsection{Confidence interval}

The problem of defining a confidence interval comes from the fact
that we do not know how to describe the statistics of errors in the
data. The Poisson noise and random errors in the
photometry do not explain the differences between independent
observations of close fields on the sky (see fig~1). A third
source of error may come from
systematic errors in the photometry. An attempt has been made to
compute the statistics of error from the comparison of data sets
towards the pole and the anticentre. It appears that neither the
number of stars in adjacent bins nor the gradient between star numbers
in adjacent bins can be used to make an estimator of error on the data.
However we can estimate the likelihood for two data sets to come
from the same true star distributions.
Using Stobie \& Ishida and Soubiran data we obtain a
likelihood of -227 for the bins in common, which corresponds to about
15 sigmas difference (Poisson noise). The
comparison between SI and GRH gives -103 (about 13
sigmas).

We may estimate the confidence interval of our fit by computing the
Poisson noise using several simulations of similar models. The
likelihood of two realizations of the same model,
differing just by the Poisson statistics, gives the value to add to
the maximum likelihood to get the confidence level.

\section{Thick disc characteristics}

The distributions in the (V, B-V) plane only has been used to determine
the structural parameters of the thick disc, because they are more
numerous and generally are deeper than U-B data. We
compare the number of stars in (V, B-V) bins obtained by model
simulations with data. The thick disc is modelled using the density law
given in Eq.~4.

\begin{equation}
\rho \propto \left\{\begin{array}{ll}
\exp{(-\frac{R-R_{\small
\sun}}{h_{R}})}*(1-\frac{1/h_{z}}{x_l*(2.+x_l/h_{z})}*z^{2}) &
\mbox{ if } z\leq x_l \\
\exp{(-\frac{R-R_{\small \sun}}{h_{R}})}*\exp({-\frac{z}{h_{z}})} &
\mbox{ if } z> x_l \\
\end{array}
\right.
\end{equation}

This density law is radially exponential and vertically exponential
for large z.
Three parameters define the density
along the z axis : $h_{z}$, the scale height, $\rho_{0}$ the local
density and $x_l$ the distance above the plane where the density law
becomes exponential. This third parameter is fixed by continuity of
$\rho(z)$ and its derivative. It allows the derivability of
the density law in the plane and to
accurately fit the density law derived from the potential using the Boltzmann
equation (see Bienaym\'e et al, 1987). Hence the parameter $x_l$ varies
with the choice of scale height and local density following the
potential. Its variation can be approximated by Eq.~5:
\begin{eqnarray}
x_l & = & 1358.6-1.35*nh+2.335 \time 10^{-4}*nh^{2} \nonumber \\
    &   & - nr*(8.1775\time 10^{-1} + 5.817*10^{-3}*nh)
\end{eqnarray}

\noindent where \[ nh = \frac{h_{z}}{1. pc} \]
\noindent and \[
nr = \frac{\rho_{0}}{1.22*10^{-3} stars~ pc^{-3}}
\]

\subsection{Scale height and local density}

The maximum likelihood method was first applied to the data towards
the galactic poles in order to measure the scale height and local
normalization independently from the scale length. The data were
binned generally by 0.1 magnitude in B-V and 1 magnitude in V, but the
data at magnitude fainter than 20 on small fields were binned by 2
magnitudes to get a sufficient number of stars per bin. We restrict
our analysis to the domain defined by V$>$13,
0.4$<$B-V$<$(B-V)$_{max}$ where (B-V)$_{max}$ depends on the limiting
magnitude of the sample. We use a (B-V)$_{max}$ of 1.0 for Schmidt
data, 0.8 for Friel \& Cudworth data (limited to magnitude 15), 1.6
for deep counts to magnitude 22. This selection allows to avoid the
region of the magnitude-colour diagram where there are few thick disc
stars in regards to thin disc stars. It reduces the contamination by
disc stars.

A grid of models has been computed with scale heights varying between
480 and 1500 pc and local density between 1\% to 15\% the one of the
disc.

When analyzing likelihood results field by field it appears that small
fields cannot constrain the free parameters : On a statistical basis
all values of scale height and normalization are compatible with the
data. This is true for Fenkart \& Esin-Yilmaz and Spaenhauer data
and also for Ratnatunga data because no real constraints on the thick
disc can be obtained at magnitude brighter than 14. At such
magnitudes thick disc stars are mainly subgiants and giants, they
are too scarce, even in 20 square degrees. But for fields
obtained from deep Schmidt plates the statistics is good and the
simulations are sensitive to the thick disc model. Figure~4 shows
contours of equal likelihood obtained for different values of scale
height and local density. Only the contours which are at less than 1
sigma are shown. The maximum likelihood obtained for each field and
the corresponding Poisson noise is given in table~1.

\begin{figure*}
\picplace{1 cm}
\caption[]{Contours of equal likelihood as a function of local density
and scale height of the thick disc population towards the galactic poles.
a) All fields; b) Stobie \& Ishida; c) Gilmore, Reid \&
Hewett; d) Koo \& Kron; e) Majewski. The contours are at 1, 2 and 3
sigmas.}
\end{figure*}

\begin{table}
\caption[]{Maximum likelihood values for each data set used in the
determination of the scale height h$_z$ and normalization $\rho_0$ of the thick
disc towards the galactic pole. L$_{\em Poiss}$ is the likelihood
computed between two realizations from the same model.}
\begin{flushleft}
\begin{tabular}{lllll}
\hline
Data set  &  h$_z$ & $\rho_0$ & L$_{max}$  & L$_{\em Poiss}$ \\
 & (pc) & (\% of disc) \\
\hline
Stobie \& Ishida & 750. & 6.1 & -113.6 & -32.2 \\
Gilmore et. al. & 800. & 4.9 & -107.5 & -6.5\\
Majewski & 700. & 6.6 & -59.0 & -2.0\\
Koo \& Kron & 840. & 6.3 & -10.2. & -0.5\\
All & 760. & 5.6 & -299. & -41.2\\
\hline
\end{tabular}
\end{flushleft}
\end{table}

Koo \& Kron data are of little use because the number of stars is too
small to give a significant constraint. There is a very good agreement
between data from Stobie \& Ishida, Gilmore et al., and Majewski. They
all together give a scale height of 760 $\pm$ 50 pc with a local density of
$5.6 \pm 0.8 $ \% of the disc density.

\subsection{Scale length}

The scale length can be derived from data at intermediate
latitudes (GCIL, Aquarius, Serpens, SA54, SA94, SA68, SA51, ACIL). The
mean galactocentric distances of thick disc stars in the different
fields studied range between 7 kpc (for GCIL and Aquarius) and 9.5
(for SA54 and ACIL) up to 11 kpc for SA51 (but for very few stars),
assuming R$_{\sun}$=8.5 kpc.  As in the analysis of the scale height
small fields do not give strong constraints on the scale length.

In the case the intermediate latitude fields would give different
result for the local density than the pole fields we solve for both parameters,
scale length and local normalization (in the range of values allowed
by the test towards the pole) for each field and for all fields
together. We assume a scale height of 760 pc, as found in the test
towards the pole. Figure~5 shows the likelihood
obtained for each data set as a function of the scale
length and local density of the thick disc. The three curves are at
1,2 and 3 sigmas respectively.
The centre field constrains the scale length to be quite small (less
than 2.5) if the
local density is small (less than 5\%) while the anticentre field
gives a looser constraint for a larger scale length if the local density
is small. The antirotation field only constrains the local density and
obviously not the scale length. The resulting
local density (5.6\% $\pm$ 0.6) is in perfect agreement with the value
found using the pole fields, and the scale length is 2.8
$\pm_{0.5}^{0.8}$.

{\it We conclude that the thick disc population can be described by a
radially exponential density law with a scale length of 2.8 kpc, a modified
exponential perpendicular to the plane with a scale height of 760 pc
and a local density of 5.6\% of the disc.}

Compared to the thin disc scale length, our result indicates a value
of the same order. Robin et al. (1992) found from star counts towards
the anticenter in the plane, that the old disc scale length is 2.5
$\pm 0.3$ kpc. Fux \& Martinet (1994) found the same value from
stellar kinematics in the solar neighbourhood. From COBE data, a 3 kpc
projected scale length has been estimated, a result well compatible
with these studies.  Concerning the thick disc scale length few
reliable measurements are available. In Ojha et al. (1995, in
preparation) the data
used in the present investigation have been pre-analyzed. From a
classical estimation of the density ratios between the centre and the
anticentre a scale length in the range 2 to 4 kpc has been obtained, the
scale height is 760 $\pm$ 50 pc and the thick disc over thin disc density
ratio 7.4 $^{+2.5}_{-1.5}$ \%. Considering the larger
uncertainties in the method involving smaller samples their result is
well in agreement with ours.

Reid \& Majewski claimed from an analysis of Majewski data that the
thick disc has a scale height larger than 1400 pc. Our study
favours, from the same data, a scale height of 760 pc. The reason is
found in the different models of disc used in both analysis. In RM
models the discs make the major part of stars at B-V$>$1.4 at
V$\sim$21, while in the Besan\c{c}on model one get a mixture of disc
and thick disc.  The disc has been carefully adjusted to data from
magnitude 5 (Hipparcos Input Catalogue) up to 22 (Prime focus plates
and CCDs), and Schmidt data at intermediates magnitude (Haywood, 1994,
Haywood et al., 1995 in preparation), while this is not the case for
RM model. Finally, our result is reinforced by the consistency
between results at high and intermediate latitudes in numerous
fields.

In order to estimate the influence of the thin disc model on the
adjusted thick disc, we attempted to use a self consistent thin disc
model computed with a maximum velocity dispersion of 18 km/s in place
of 21 used in simulations presented above. The adjusted thick disc
characteristics are very closed to previous ones, giving a slightly
higher local density (to compensate the smaller thin disc) but the
difference is at less than 1 sigma. Moreover the likelihood obtained
with this thinner disc is slightly smaller than with a thin disc of
$\sigma_{W max}$ = 21 km/s.

\subsection{Luminosity function}

In this study we assume that the luminosity function of the thick disc
is similar to the one of 47Tuc. Can we constrain this function using
these data and would the choice of the LF influence the result on the
density ?

The shape of the luminosity function depends mainly on three
parameters : the age, the slope of the Initial Mass Function and the
metallicity, assuming that the thick disc formed on a small time
scale relative to the age of the Galaxy. The metallicity is derived
from colour informations (sect. 5). The age parameter fixes the
absolute magnitude of the turnoff, where the LF is the
steepest. A difference of 4 Gyr gives a shift
of 0.4 magnitude only. Turnoff
stars of the thick disc appear in star counts at high latitude at
about 14-15 V magnitude. Available counts are still not sufficient
to constrain such a small absolute magnitude shift.

Luminosity functions computed by Bergbush \& Vandenberg (1992) vary
most significantly at the level of faint stars for different IMF (M$_V>9$).
We used their models to test two hypothesis for the IMF slope (x=0.5
and x=1.0). The maximum likelihood was computed for two apparent
magnitude ranges : 15-17 and 17-22. The ratio of the number of stars
in both ranges gives a constrain independent from
the scale height. The best agreement between model and data is
obtained with a
slope of x=1 but a slope of 0.5 cannot be rules out at a 1 sigma level.

\begin{figure*}
\picplace{1 cm}
\caption[]{Likelihood as a function of the scale length and local density
computed from fields : a) GCIL (4,47); b) ACIL (169,47); c) ARIL (278,47);
d) All fields together.}

\end{figure*}

\subsection{Abundances}

Colour distributions are also sensitive to the metal abundances of the
populations. While U-B is more sensitive to the metallicity,
B-V data tend to be more accurate. This is the reason why we
attempt to estimate the mean metallicities and gradients of the
various populations using
separately the (V,B-V) plane and the (V,U-B) plane. In the case of
U-B we restrict our analysis to 7 fields (North Galactic Pole, from
Spaenhauer (1989), from Fenkart \& Esin-Yilmaz (1985) between 10 and
18, from Koo \& Kron (1982) between 20 and 22, (l=4, b=47) from Ojha
et al., 1994a, (l=169, b=47) from Ojha et al., 1994b, SA54 from
Fenkart \& Esin-Yilmaz, 1983).

Starting with a simulated catalogue for each data set, we compute for
each star its colours assuming different metallicities. Then we
compute the resulting (V, U-B) and (V, B-V) distributions and compare
them with real data.  We apply the same maximum likelihood test as in
the previous section. Colour data are binned in steps of 0.1 magnitude
and V in steps of 1 magnitude, restricted to V$>$14 and U-B$<$1.2.

Using this method we solve for several parameters simultaneously: the mean
metallicity of the thick disc and the radial and vertical gradients of the
thin disc.

\begin{figure*}
\picplace{1 cm}
\caption[]{Likelihood obtained for different samples as a function of
the mean metallicity of the thick disc. a: Pole Schmidt data
(Stobie \& Ishida); b: Pole, other data (solid line : Spaenhauer,
dashed line: Fenkart \& Esin-Yilmaz, dotted-dashed: Koo \& Kron);
c: GCIL field (l=4, b=47, solid line) and Aquarius (37, -51, dashed line);
d: Anticentre fields (Solid line: Yamagata \& Yoshii (SA54), dashed
line: Ojha et al. (l=169, b=47), dotted-dashed: Fenkart \&
Esin-Yilmaz (SA54). In bold the range of acceptable values.}
\end{figure*}

Figure~6 shows the resulting likelihoods obtained for the thick disc
mean metallicity. Table~3 gives the values of maximum likelihood and
acceptable 1 sigma range field by field. The confidence interval is
computed following the method described in section 4. All fields
together give a mean metallicity of -0.7 $\pm$ 0.2. We find no
noticeable gradient of abundances between centre and anticentre
fields.

\begin{table}
\caption[]{Range of acceptable values for the mean metallicity of the
thick disc obtained by
maximum likelihood test, at 1 sigma level, as described in the text.}
\begin{flushleft}
\begin{tabular}{lllll}
\hline
Data set  &   L$_{max}$ & [Fe/H] & L$_{\em Poisson}$ & Range\\
\hline
Pole SI & -290. & -0.5 & -90. & [-0.4, -0.7]\\
Pole Spaenhauer & -28 & -1.0 & -5.4 & [-0.6, -1.3]\\
Pole FEY & -26 & -1.0 & -4.2 & [-0.7, -1.3] \\
Centre GCIL & -426 & -0.8 & -98 & [-0.6, -1.0]\\
Centre Aquarius & -1229 & -0.8 & -64 & [-0.6, -1.0]\\
Anticentre YY & -127 & -0.7 & -75 & [-0.5, -1.0]\\
Anticentre ACIL & -50 & -0.9 & -52 & [-0.5, -1.3]\\
Anticentre FEY & -34 & -1.1 & -4.3 & [-0.9, -1.2]\\
All & -2383 & -0.7 & -342 & [-0.6, -0.9]\\
\hline
\end{tabular}
\end{flushleft}
\end{table}

Exploring the possibility that the halo metallicity influences
the result, we found that the available data do not go deep enough in
V to get sufficient information on the metallicity of the halo and,
conversely, that its adopted value does not influence the
result about the thick disc metallicity.

An attempt has been made to measure the metallicity gradients in the
thin disc population. The radial gradient was found in the range 0 to
-0.2 dex kpc$^{-1}$, values less compelling than what can be
obtained by direct measurements of metallicities in stellar samples.
Concerning the vertical gradient, values obtained using different
samples are contradictory. SI data towards the pole are in favour of a
small gradient of -0.25 dex kpc$^{-1}$, YY data in SA54 give
a high value of -0.65, while FEY favour -0.15 dex kpc$^{-1}$. Ojha
(1994a) data in ACIL field are nearly insensitive to this parameter. We
conclude that this effect is at the limit of the accuracy of the
present data.

\section{Thick disc formation}

Two families of scenarii are advocated to explain the formation of the
thick disc. The first family features `pre-thin disc' or `top-down' models
where the thick disc forms through a dissipational collapse, after
the halo formation and before the thin disc has completely
collapsed. {In top down scenarii thick disc stars form during late
phases of the collapse, which means either a slow collapse or a high
star formation rate during these phases. Slow down collapse regimes
generate chemical and kinematical gradients which would be clear
signatures of such a process. In contrast, top-down models with
enhanced star formation offer a possible gradient-free
mechanism. However, playing freely with ad-hoc star formation rate, in
the absence of an identified enhancement mechanism looks quite
arbitrary. We show elsewhere (Haywood et al., 1995) that along the
whole thin disc lifetime the star formation rate has been stable or
slightly increasing. Along the same lines, top-down scenarii hardly
generate a discontinuity between thick disc and thin disc, unless some
ad-hoc mechanism switches off star formation during a transition phase.}

The second type of models are called `post-thin disc'. They resort to
formation of the thick disc after the gas has completely collapsed into
a thin disc. Possible physical processes are: 1) Secular kinematic
diffusion of thin disc stars. In this case the thick disc is a
prolongation of the thin disc, the gradients visible in the thin disc
should appear in the thick disc.
2) Violent thin disc heating by
the accretion of a satellite galaxy (The
required event(s) must not occur too late in the disc life time so that the
gas can cool again and form stars in the long-lasting thin galactic disc.
3) Accretion of a satellite galaxy without heating of the thin
disc (thick disc is formed of the debris of the satellite(s)). These last
two scenarii can be combined.

The violent bottom up scenario leaves two important observational
signatures. {First the thick disc is a separate population distinct from
the thin disc and the halo. Second no gradient can be generated in the
thick disc by the event, although a pre-existing gradient may survive
the merger.}
One can compare the thick disc
observed characteristics with these different signatures.

\begin{enumerate}

\item {\it There is a change of slope in the density law between thin
and thick disc}. The scale heights obtained for the thin and thick disc are
strongly different. Haywood (1994) shows, from an analysis of numerous
star counts towards the pole using his self-consistent evolutionary
model, that the thin disc scale
height does not exceed 250 pc (K dwarfs). This result is confirmed by
Ojha (1994)
from a standard analysis of stellar densities towards the pole. A similar
result has also been obtained by Kroupa (1992): A scale height of 270 pc
for stars of M$_V>6$.
In the present paper the thick disc is shown to have a scale height of
760 $\pm$ 50 pc using a synthesis approach free of bias. Same value is
also obtained from a classical analysis of star counts by Ojha et
al. (1994ab) and Soubiran (1992).

The three basic quantities characterizing the thick disc density
distribution (local density, scale height, scale length) are tightly
and consistently constrained by the global analysis over different
fields at various latitudes and longitudes, a consistency which had
never been obtained by previous studies.

Hence density laws show a true discontinuity between thin disc and thick disc
populations.

\item {\it The thick disc is kinematically distinct from the thin disc
and from the halo; it shows no kinematic gradients}.
Kinematic data from Ojha et al. (1994ab) use a SEM
algorithm to estimate the number of Gaussian components
present in proper motion data towards 3 directions
at intermediate latitude. {In the SEM analysis,
discrete Gaussian kinematic components are considered. The data do
support this hypothesis, although more complex distribution functions
might be valid.}
In all cases they obtain a clear separation between
a thin and a thick population. For example, towards the
anticentre (Ojha et al, 1994) :

\[  \sigma_{\frac{U+W}{\sqrt{2}}})_{thick} -
\sigma_{\frac{U+W}{\sqrt{2}}})_{thin} = 30 \pm 3 km s^{-1} \]
\[  \sigma_{V})_{thick} - \sigma_{V})_{thin} = 30 \pm 3 km s^{-1} \]
\[  {\em Asymmetric~drift} :~<V>_{thick} - <V>_{thin} = 45 \pm 4 km s^{-1} \]

All recent studies of the asymmetric drift of the thick disc converge
to small values (about 40 km s$^{-1}$, Ojha, 1994). Most authors
do not find any vertical gradient in any of the three fields. However
Majewski (1992) invokes
such gradient and, on this basis, advocates a dissipational collapse for
the thick disc. His figure 6 shows a number of results from different
samples of supposed thick disc stars. However most data are
selected sample based on the metallicity and proper motions. These
samples are generally contaminated by disc stars at low z and by halo
stars at high z. It have been shown by Ojha et al. (1994ab) and
by Soubiran (1992) that, when one
separates the populations on a statistical basis by multi-Gaussian
fitting without a priori informations on them, one can separate a thin
disc from a thick disc and the asymmetric drift shows no gradient with
z inside the thick disc population itself. Moreover no velocity
dispersion gradient is observed inside the thick disc (Ojha et al, 1994ab).

On the other hand there is a large gap of circular velocity between
the thick disc (180 km s$^{-1}$) and the halo (0 $\pm$ 40 km
s$^{-1}$), proving the
disentangling of these two populations.

Hence the kinematics of the thick disc shows that it is a population
distinct from the thin disc and from the halo and that no vertical
gradient is present. Both arguments are
incompatible with a 'top-down' scenario of formation for the thick
disc.

\item {\it Thick disc stars show no vertical abundance gradient}
However the U-B star counts may be suspected of systematic errors we
find that the thick disc has a mean metallicity [Fe/H] of -0.7 dex and
find no evidence for a vertical gradient of this quantity. This result
is confirmed by observations of Gilmore \& Wyse (1995) of  F/G thick
disc stars. They also obtain a mean [Fe/H] of $-0.7$ dex and no metallicity
gradient in the distance range 500 to 3000 pc.
{This result argues against a 'top-down' model of thick disc with a
slow dissipational collapse, although this argument cannot
throw away a top-down model with fast collapse and a very high star
formation rate. On the other hand, QHF simulations of the
merging scenario show that the event does not alter substantially
the pre-existing gradient, if any. Observations of no abundance
gradient in the thick disc imply no previous gradient in the thin
disc. This is consistent with our observations that the thin disc
actually has no metallicity gradient, apart from the one produced by
orbit diffusions.}

\end{enumerate}

All observed characteristics favour a post-thin disc formation
for the thick disc. If this is caused by a secular heating of the thin
disc one should see a continuity with the thin disc. The large
$\sigma_W$ of the thick disc implies a heating process
particularly efficient and on a short period (the $\sigma_W$ of the
thick disc is about 42 km s$^{-1}$ while the disc does not exceed
21~km~s$^{-1}$) (Haywood, 1994).
Hence this scenario is rejected by the observations.
We stay with two scenarii related to the merging of a satellite galaxy
with the Milky Way. Statler (1988) studied the possibilities that a
merging satellite leaves a part of its material in the Galaxy. He
showed that the residuals should form a peanut shaped bulge but also
form a kind of thick layer. However he did not study the
impact on the thin disc of this accretion. This scenario is only valid
if the accretion is made of small satellites in order to conserve a
thin disc. Quinn, Hernquist \& Fullagar (1993,
hereafter QHF) analyze a scenario of satellite
accretion by a spiral galaxy. A thick disc may be produced by the dynamical
heating of the thin disc. The resulting thick
disc has typical characteristics that may be compared with our new
measurements of the thick disc of the Galaxy. No vertical kinematic gradients
should be present, an abundance gradient may be present only if it was
present before the event. In the example given by QHF the resulting thick disc
has a scale height of 750 pc in the solar neighbourhood with
projected values ranging from 850 to 1200 pc in the galactocentric
distance range $7<R<10.5$. Although we
do not see any rise of it with radius, this is not excluded because
few stars are reached at large enough distances.
For the same reason the rise of velocity dispersion in
the external part of the Galaxy, implied by the QHF model, is
significant at too large distances to be seen in the Ojha et al. data.

{We conclude that the characteristics of the thick disc as
measured by photometric and proper motion star counts show a
population quite distinct from the thin disc and well in agreement
with a disruptive accretion scenario. These characteristics do
exclude thick disc formation via slow dissipative collapse prior to
the formation of the (thin) disc. Other interpretations can be built,
based on abrupt {\it ad hoc} variations of the star formation rate,
although tuning such interpretations to combined kinematic and
photometric constraints looks rather arbitrary.}

\acknowledgements{This study has been partially made under the support
of the IFCPAR-CEFIPRA, Indo-French Center for the Promotion of
Advanced Research -- Centre Franco-Indien pour la promotion de la
Recherche Avanc\'ee. We thank Andreas Spaenhauer for allowing us to use
his data in advance of publication.}

{}
\end{document}